\pdfoutput=1
\documentclass[11pt]{article}

\usepackage[]{ACL2023}

\usepackage{times}
\usepackage{latexsym}

\usepackage[T1]{fontenc}
\usepackage[utf8]{inputenc}

\usepackage{microtype}

\usepackage{inconsolata}

\usepackage{booktabs}
\usepackage{multicol}
\usepackage{multirow}
\usepackage{makecell}
\usepackage{cleveref}
\usepackage{graphicx}
\usepackage{float}

\usepackage{tikz}
\pgfdeclarelayer{background}
\pgfdeclarelayer{main}
\pgfsetlayers{background,main}
\usetikzlibrary{calc}

\newcommand*{\minWidth}{15}
\newcommand*{\maxValue}{60}

\newcommand{\makeSimpleBar}[3]{%
    \hspace{-9pt}
  \tikz[baseline]{
    \node[anchor=base,text width=\minWidth,align=#2,inner sep=0pt,inner xsep=\tabcolsep,outer sep=0pt] (n) {\strut{#1}};
    \begin{pgfonlayer}{background}
        {
            \edef\color{#3} 
            \pgfmathparse{abs(#1/(\maxValue))}
            \edef\contents{{\pgfmathresult}}
            \fill[font=\boldmath,color=\color] ($(n.north west)$) rectangle ($(n.south west)!{{\contents}}!(n.south east)$);
        }
    \end{pgfonlayer}
  }
}

\newcommand{\pbar}[1]{\makeSimpleBar{#1}{center}{green!25}}
\newcommand{\rbar}[1]{\makeSimpleBar{#1}{center}{cyan!25}}
\newcommand{\fbar}[1]{\makeSimpleBar{#1}{center}{orange!25}}

\usepackage{xcolor}

\title{On The Importance of Reasoning for Context Retrieval in Repository-Level Code Editing}

\author{Alexander Kovrigin$^*$ \\
  JetBrains Research Germany \\
  \texttt{alexander.kovrigin@jetbrains.com} \\\And
  Aleksandra Eliseeva$^*$ \\
  JetBrains Research Serbia \\
  \texttt{alexandra.eliseeva@jetbrains.com} \\\AND
  Yaroslav Zharov \\
  JetBrains Research Germany \\
  \texttt{yaroslav.zharov@jetbrains.com} \\\And
  Timofey Bryksin \\
  JetBrains Research Cyprus \\
  \texttt{timofey.bryksin@jetbrains.com} \\}

\begin{document}
\maketitle
\def\thefootnote{*}\footnotetext{These authors contributed equally to this work}\def\thefootnote{\arabic{footnote}}
\begin{abstract}
Recent advancements in code-fluent Large Language Models (LLMs) enabled the research on repository-level code editing. In such tasks, the model navigates and modifies the entire codebase of a project according to request. Hence, such tasks require efficient \textit{context retrieval}, \emph{i.e.}, navigating vast codebases to gather relevant context. Despite the recognized importance of context retrieval, existing studies tend to approach repository-level coding tasks in an end-to-end manner, rendering the impact of individual components within these complicated systems unclear.
In this work, we decouple the task of context retrieval from the other components of the repository-level code editing pipelines.
We lay the groundwork to define the strengths and weaknesses of this component and the role that reasoning plays in it by conducting experiments that focus solely on context retrieval\footnote{The code is available on GitHub \url{https://github.com/JetBrains-Research/ai-agents-code-editing}}.
We conclude that while the reasoning helps to improve the precision of the gathered context, it still lacks the ability to identify its sufficiency.
We also outline the ultimate role of the specialized tools in the process of context gathering.

\end{abstract}

\section{Introduction}

The advances in large language models (LLMs) inadvertently drew the attention of researchers and practitioners to their possible real-world applications~\cite{llm-survey}. In particular, LLMs have shown outstanding capabilities in the software engineering domain, enabling the rise of programming assistants and allowing the research community to tackle complicated tasks close to a software engineer's everyday workflow~\cite{llm4se-survey}. 

Recently, there has been no shortage of works on repository-level coding tasks, such as code completion~\cite{repocoder, repohyper, repoformer}, code editing~\cite{codeplan, swebench, sweagent, autocoderover}, and other~\cite{rrr, codeagent, repoagent, chatdev, agentfl}. Such tasks are highly practical, but they imply mimicking software engineer's daily work, including working with large codebases spanning thousands of lines of code.

Current findings show that \textit{context retrieval}---the process of navigating through the codebase to find the relevant code---remains one of the main challenges of the repository-level coding tasks and allows to boost the end performance significantly~\cite{swebench, repohyper}. For instance, on SWE-bench, the renowned benchmark for resolving real-world GitHub issues, providing ground truth context instead of using a simple Retrieval-Augmented Generation (RAG) with a BM25-based system~\cite{bm25} leads to $144.9\%$ increase in the number of correctly resolved issues for the best-performing model, Claude 2~\cite{swebench}.

While the research community agrees on the importance of context retrieval for repository-level coding tasks, the experiments are often conducted in an end-to-end fashion, making the impact of each individual component ambiguous. For instance, AutoCodeRover~\cite{autocoderover} proposes non-trivial improvements to the context retrieval step by incorporating code structure-aware tools and reasoning techniques like self-reflection~\cite{selfrefine}. However, the context retrieval strategy is introduced as an end-to-end approach, making the precise impact of each individual component unclear. Furthermore, many other works on repository-level coding tasks tackle them with an LLM-based agent~\cite{sweagent, repoagent, codeagent} or multiple LLM-based agents~\cite{agentfl, chatdev}, where codebase navigation becomes but one of many tools available to the agent.

Given the importance of context retrieval, we argue that information about the performance of different approaches to context gathering is important on its own. Hence, we embark on a journey to study context retrieval strategies for repository-level coding tasks. 

% Specifically, we select the context retrieval strategy from AutoCodeRover~\cite{autocoderover} as the most sophisticated one as of the date and the BM25~\cite{bm25} retriever as the simple baseline. Inspired by the research on RAG in a natural language domain, which shows the promise of agentic approach~\cite{react} and reasoning techniques~\cite{self-rag, crag}, we also consider \note{six} strategies building off similar ideas. We conduct experiments on two repository-level code editing benchmarks, SWE-bench~\cite{swebench} and LCA Code Editing~\cite{lca-code-editing}. 

% TODO: add findings

\section{Related Works}

% \subsection{Context Retrieval in Repository-level Coding Tasks} 

Context retrieval is an essential step for repository-level coding tasks tackled in multiple previous works.

Standard approaches from the natural language processing domain are widely employed for context retrieval. For example, SWE-bench~\cite{swebench} utilizes a standard Retrieval-Augmented Generation (RAG) approach with a BM25 retriever~\cite{bm25}. 
In the case of the classical RAG, we first make a request to the knowledge base---codebase in our case---and then add the result of such request to the prompt of the model to condition the prediction of the model on the retrieved knowledge~\citet{rag-survey}.

RepoCoder~\cite{repocoder} also uses RAG but performs multiple iterations to enhance the performance.
In this case, the iterations are done without reasoning. 
They generate a chunk of code iteratively and use it as input to search for possible related chunks in the codebase.

As the next step to increase the retrieval performance, SWE-agent~\cite{sweagent} introduces a ReAct-style reasoning~\cite{react} to the process, equipping the reasoning model with tools for navigating through files and directories.
ReAct-based algorithms perform the retrieval in a series of generations. 
The model is prompted to consider the newly acquired information's usefulness, decide if it should be added to the context, and then generate a new search request.

Finally, the next reasoning improvement in this line of research is a separate reasoning step of Self-Reflection introduced by~\citet{selfrefine}.
In this step the model is separately prompted to consider if the currently collected context is enough for the task at hand.
This step is used, for example, in the AutoCodeRover~\cite{autocoderover} approach.

Another branch of development in the repository level code-retrieval is the usage of code-specific tools.
One common method is to use a graph representation of the repository's codebase, where nodes represent code entities and edges denote their relations. 
Such graphs naturally facilitate context retrieval for coding tasks. 
For instance, CodePlan~\cite{codeplan} builds the context based on the static dependencies defined in the graph, while RepoHyper~\cite{repohyper} captures both static and more implicit relations by first retrieving a set of relevant nodes via semantic representations and then further extending it via graph search algorithms.

%For instance, RepoHyper~\cite{repohyper} retrieves the context from a repository graph, while CodePlan~\cite{codeplan} leverages static dependencies within the graph to gather context for code editing and to incrementally update all the code locations affected by previous edits.

Combining reasoning with the specialized tools, AutoCodeRover~\cite{autocoderover}, RRR~\cite{rrr}, and CodeAgent~\cite{codeagent} equip an LLM-based agent with a set of code structure-aware tools to navigate through the codebase.

% Other methods to utilize code structure for repository-level tasks include, for instance, enabling the RAG system to decide if the retrieval is necessary~\cite{repoformer}. 

% \subsection{Reasoning in Retrieval-Augmented Generation}

% Retrieval-Augmented Generation itself is a vast research area covered, \emph{e.g.}, in the survey from~\citet{rag-survey}. In particular, numerous works have been done on incorporating reasoning techniques within the RAG systems. Self-RAG~\cite{self-rag} and CRAG~\cite{crag} demonstrate the advantages of including self-reflection (a reasoning technique introduced by \citet{selfrefine}) on various stages of RAG: for instance, to judge whether the retrieval is necessary in the first place or to assess the relevance of each retrieved document. Another frequent paradigm is to query external retriever iteratively~\cite{react}%DSP?~\cite{dsp}
% , as a natural extension of Chain-of-Thought, the common reasoning paradigm to decompose tasks~\cite{cot}.
Unlike all the works presented in this section, we aim to cover context retrieval specifically for repository-level code editing. On the other hand, we consider the retrieval decoupled from the other parts of the pipeline.

\section{Experiments \& Results}

\subsection{Models}
Throughout all the experiments, we use a proprietary LLM GPT-3.5 Turbo (\texttt{gpt-3.5-turbo-16k}) through the official OpenAI API.\footnote{\url{https://platform.openai.com/docs/models/gpt-3-5-turbo}} This model is often used as a go-to closed-source model, faster and cheaper than more advanced LLMs while still offering competitive capabilities~\cite{chatbot-arena} and a sufficiently large context size of 16k tokens. We aim to extend the model list in the future.

\subsection{Datasets}
We select two repository-level code editing datasets with different context complexity to test the performance in varied environments. 
\textit{SWE-bench}~\cite{swebench} is a renowned benchmark consisting of texts of real-world issues as inputs and the corresponding patches as targets. It contains 2,294 data points from 12 GitHub repositories. In this work, we consider SWE-bench Lite, a smaller subset of SWE-bench with 300 issues across 11 Python repositories. 
\textit{LCA Code Editing}~\cite{lca-code-editing} is another repository-level code editing dataset consisting of curated commit messages that serve as natural language instructions and corresponding code changes as the target. It contains 119 data points from 39 GitHub repositories. One distinguishing feature of this dataset is its focus on large-scale changes---the average number of lines in the gold patches for LCA is almost $8$ times larger than for SWE-bench Lite---which makes context retrieval naturally harder. We provide the statistics on the average context length in both datasets in \Cref{tab:len-stats}.

\begin{table}
\centering
\begin{tabular}{c c cc cc}
\toprule
& & SWE-Lite & LCA\\
\midrule
Prompt & \#Tokens & 444 & 39\\
\midrule
\multirow{3}{*}{Patch} 
& \#Tokens & 120 & 1,237\\
& \#Lines & 25.9 & 206.5\\
& \#Files & 1 & 4.3\\
\midrule
\multirow{3}{*}{Code} 
& \#Tokens & 3.7M & 1.4M\\
& \#Lines & 360K & 186K\\
& \#Files & 1,580 & 1,055\\
\bottomrule
\end{tabular}
\caption{Mean context length in the considered datasets. Tokens are obtained via the GPT-3.5 Turbo tokenizer. SWE-Lite stands for SWE-bench Lite.}\label{tab:len-stats}
\end{table}

\subsection{Context Retrieval Strategies}
%To compare various context retrieval strategies, we select a range from the BM25~\cite{bm25} retriever as the simple baseline to the AutoCodeRover (ACR)~\cite{autocoderover} as the most sophisticated one as of the date. 

We select a wide range of context retrieval strategies. %A brief introduction to each follows hereinafter.
We include a simple baseline of making a single request to the standard \textit{BM25}~\cite{bm25} retriever, which was previously applied to repository-level code editing by~\citet{swebench}. 
In our setup, we select several top retrieved documents to ensure a context size of at least 500 tokens.

The rest of the context retrieval strategies are based on \textit{ReAct}-style agents~\cite{react}, \emph{i.e.}, they iteratively query LLM in a loop, interleaving reasoning and acting. 
%\input{tables/02-retrieval-strategies}
%\Cref{tab:context-retrieval-approaches} summarizes the key features of all agentic context retrieval strategies considered in this work. 
We vary two components: external \textit{tools} that the agent is equipped with and the \textit{stopping criteria} for the agent loop.

Regarding the toolset, we consider two options: a simple BM25 retriever and a set of code structure-aware tools proposed in AutoCodeRover (ACR)~\cite{autocoderover}.

We investigate three versions of stop conditions, each progressively enhancing the model's reasoning capabilities.
The simplest stopping criterion is \textit{Context Length (CL)}, which keeps iterating until the size of the gathered context achieves at least $500$ tokens.
The second stopping criterion is \textit{Tool Call (TC)}, which resumes iterations until the first LLM output without a tool call. This approach is common in existing agentic frameworks, \emph{e.g.}, LangChain~\cite{langchain}. 
The third stopping criterion, \textit{Self-Reflection (SR)}, extends \textit{TC} by explicitly querying the LLM to assess whether the current context is sufficient or if further execution is needed.

Finally, we consider \textit{AutoCodeRover (ACR)}, the most sophisticated context retrieval strategy as of the date. ACR combines the most advanced reasoning from the above with specialized tools.
However, as it follows more complicated execution logic than the agents we use and contains advanced prompts, we consider it to be one step further on the axis of reasoning.

To summarize, we vary the complexity of the tested approaches along two axes.
Along the tools axis, we have two possible positions: BM25 and ACR tools.
Along the reasoning axis, we have 5 positions: baseline, CL, TC, SR, and ACR, listed by increasing their reasoning complexity\footnote{Note that the baseline is evaluated only with BM25, and ACR is evaluated only with ACR tools, as these tools are integral to their respective definitions.}.

\subsection{Metrics} 
\citet{swebench, repohyper} show that the quality of context retrieval directly affects the end results on the downstream tasks. Thus, we focus on evaluating context retrieval as a standalone component and leave exploring the downstream performance to future works. We consider the standard localization metrics: \textit{Precision}, \textit{Recall}, and \textit{F1}. 
The recall is an important metric because without retrieving the correct part of the codebase, the model will not be able to modify it.
Precision is an important metric because the models tend to perform worse given irrelevant information as part of the prompt.
We use F1 as a classical metric that unifies both of them, though we consider proper mapping of relative importance of precision and recall for the downstream performance a task for future research.

We report the localization metrics on two scopes of varying granularity: on the level of \textit{files} and on the level of specific code \textit{entities} (i.e., classes and functions). %, and on the level of specific code \textit{lines}. 
For each level, we compare the retrieved context with the affected elements indicated by the ground truth patch.

\subsection{Results \& Discussion}

% \note{TODO: add more concrete numbers to the discussion, compress it and improve it overall}
\begin{table*}[h]
\centering
% \footnotesize % makes all fonts smaller inside the table
% \setlength{\tabcolsep}{1pt} % defines horizontal spacings in the table
% \def\arraystretch{0.5} % defines the vertical spacings in the table
\begin{tabular}{ll|lll|lll|c}
\toprule
% & \multicolumn{3}{c}{\textbf{File-level}} & \multicolumn{3}{c}{\textbf{Entity-level}} & \multicolumn{3}{c}{\textbf{Line-level}} 
& & \multicolumn{3}{c|}{\textbf{File-level}} & \multicolumn{3}{c|}{\textbf{Entity-level}} & \multirow{2}{*}{\textbf{Avg. CL}}\\
\cmidrule(r){3-5} \cmidrule(l){6-8}
& & P & R & F1 & P & R & F1 & \\
\midrule
\multicolumn{9}{c}{\textbf{SWE-Bench Lite}}\\
\midrule
\multirow{4}{*}{\rotatebox[origin=c]{90}{\textbf{BM25}}} 
& Baseline & \pbar{9.2} & \rbar{29.6} & \fbar{13.5} & \pbar{4.8} & \rbar{17.6} & \fbar{7.2} & 493\\
& ReAct + CL & \pbar{11.4} & \textbf{\rbar{52.0}} & \fbar{17.4} & \pbar{4.7} & \textbf{\rbar{24.3}} & \fbar{7.4} & 950\\
& ReAct + TC & \pbar{22.3} & \rbar{35.8} & \fbar{26.2} & \pbar{10.4} & \rbar{15.8} & \fbar{11.5} & 278\\
& ReAct + SR & \textbf{\pbar{25.6}} & \rbar{40.3} & \textbf{\fbar{29.6}} & \textbf{\pbar{14.2}} & \rbar{18.6} & \textbf{\fbar{14.3}} & 246\\
\midrule
\multirow{4}{*}{\rotatebox[origin=c]{90}{\textbf{ACR Tools}}} 
& ReAct + CL & \pbar{29.8} & \rbar{61.1} & \fbar{36.8} & \pbar{12.6} & \textbf{\rbar{42.2}} & \fbar{17.0} & 1303\\
& ReAct + TC & \pbar{34.9} & \rbar{45.6} & \fbar{37.7} & \pbar{20.1} & \rbar{23.9} & \fbar{19.1} & 402\\
& ReAct + SR & \pbar{42.0} & \rbar{50.6} & \fbar{44.4} & \pbar{25.8} & \rbar{30.6} & \fbar{25.4} & 306\\
& ACR (custom) & \textbf{\pbar{55.2}} & \textbf{\rbar{61.6}} & \textbf{\fbar{56.8}} & \textbf{\pbar{30.5}} & \rbar{34.0} & \textbf{\fbar{28.9}} & 763\\
\midrule
\multicolumn{9}{c}{\textbf{LCA}}\\
\midrule

\multirow{4}{*}{\rotatebox[origin=c]{90}{\textbf{BM25}}} 
& Baseline & \pbar{18.8} & \rbar{25.3} & \fbar{18.9} & \pbar{12.7} & \rbar{9.6} & \fbar{9.4} & 487\\
& ReAct + CL & \pbar{22.8} & \textbf{\rbar{36.5}} & \textbf{\fbar{23.6}} & \pbar{14.4} & \textbf{\rbar{12.4}} & \textbf{\fbar{11.4}} & 846\\
& ReAct + TC & \pbar{29.6} & \rbar{22.2} & \fbar{22.4} & \textbf{\pbar{21.0}} & \rbar{7.0} & \fbar{8.9} & 196\\
& ReAct + SR & \textbf{\pbar{30.4}} & \rbar{21.2} & \fbar{21.7} & \pbar{20.7} & \rbar{7.8} & \fbar{9.4} & 198\\
\midrule
\multirow{4}{*}{\rotatebox[origin=c]{90}{\textbf{ACR Tools}}} 
& ReAct + CL & \pbar{47.3} & \textbf{\rbar{45.7}} & \fbar{39.3} & \pbar{23.7} & \textbf{\rbar{21.4}} & \fbar{19.1} & 1599\\
& ReAct + TC & \pbar{46.6} & \rbar{36.8} & \fbar{35.0} & \pbar{30.8} & \rbar{11.9} & \fbar{13.2} & 557\\
& ReAct + SR & \pbar{49.6} & \rbar{31.4} & \fbar{34.5} & \pbar{36.3} & \rbar{12.6} & \fbar{15.2} & 568\\
& ACR (custom) & \textbf{\pbar{62.5}} & \rbar{36.3} & \textbf{\fbar{39.8}} & \textbf{\pbar{42.6}} & \rbar{17.6} & \textbf{\fbar{20.2}} & 956\\
\bottomrule
\end{tabular}
\caption{(P)recision, (R)ecall, and F1 scores of different context-retrieval strategies depending on the reasoning approaches and tools used.  The best results among each set of tools, scope, and dataset are highlighted in bold. Average context length (CL) is reported in tokens obtained from the GPT-3.5 Turbo tokenizer. Color bars represent the value inside the cell to ease visual analysis.}
\label{tab:results}
\end{table*}
We report our quantitative results in~\Cref{tab:results}.
Further in this section, we present and discuss our observations driven by those results.

Our first observation is that precision is highly correlated with the increase in reasoning complexity, not with the context size in tokens.
The correlation coefficient between precision and reasoning levels is more than 0.7 for both file and entity context levels and small but positive 0.08 with the context length for the entity-level context.
We conclude that reasoning allows us to collect more context with better precision.

Our second observation is that recall mostly correlates with the context length. 
The correlation between the context length and recall is 0.5 on average, while 0.1 on average with the reasoning level.
We conclude that reasoning plays a role, but it is not enough to decide whether the context is sufficient to solve the task.

Our third observation is that giving an agent task-specific search tools grants huge performance improvements.
There are almost no cases in the \Cref{tab:results}, where any agent with specialized tools performs worse than an agent with just the search tool, without regard to their reasoning tools.

\section{Conclusion}

In this study, we evaluated the impact of individual components of the context retrieval strategies for the repository-level code-editing tasks, namely, code structure awareness and reasoning.
We conclude that reasoning plays the ultimate role in increasing precision in the retrieved context\footnote{The code is available on GitHub \url{https://github.com/JetBrains-Research/ai-agents-code-editing}}.
On the other hand, recall is mostly regulated by the length of the context, which prompts further research of reasoning approaches to estimate the sufficiency of the gathered context.
%The specialized tools are also of great importance for context generation and therefore, further research on the synergy between reasoning abilities and tool usage is needed.
%\note{Agent-computer interface, reasoning, and so on.}
Specialized tools are also of great importance for context retrieval. As noted by \citet{sweagent}, further research into Agent-Computer Interfaces---how to design the interactions between LLMs and external environments to maximize the reasoning potential---could be crucial for improving performance.
Overall, we argue that reasoning for retrieval is an important research area that should be studied rigorously.

\section*{Limitations}
This paper presents preliminary findings and is part of ongoing research. The current limitations are as follows:
\begin{itemize}
            \item This study relies on one proprietary Large Language Model (LLM), which may limit the generalizability of the results. Future work will involve evaluating multiple LLMs to enhance the robustness of the findings.
            \item Only a limited number of context retrieval approaches were explored. Expanding the range of methods in future research will provide a broader perspective on their effectiveness and applicability.
\end{itemize}

% \section*{Ethics Statement}
% Scientific work published at ACL 2023 must comply with the ACL Ethics Policy.\footnote{\url{https://www.aclweb.org/portal/content/acl-code-ethics}} We encourage all authors to include an explicit ethics statement on the broader impact of the work, or other ethical considerations after the conclusion but before the references. The ethics statement will not count toward the page limit (8 pages for long, 4 pages for short papers).

% \section*{Acknowledgements}

\bibliography{anthology,custom}
\bibliographystyle{acl_natbib}

% \appendix

% \section{Example Appendix}
% \label{sec:appendix}

% This is a section in the appendix.

\end{document}